\documentclass[preprint,5p]{elsarticle}

\makeatletter
\def\ps@pprintTitle{%
 \let\@oddhead\@empty
 \let\@evenhead\@empty
 \def\@oddfoot{\centerline{\thepage}}%
 \let\@evenfoot\@oddfoot}
\makeatother

\usepackage{graphics}
\usepackage{graphicx}
\usepackage{subfig}
\usepackage{amsmath,amssymb,esint}
\usepackage{amsthm}
\usepackage{float}   
\usepackage{caption}
\usepackage[dvipsnames]{xcolor}
\usepackage[colorlinks]{hyperref}

\newcommand{\rev}[1]{{\color{black} #1}}
\biboptions{sort&compress}

\journal{Ultrasonics Sonochemistry}

\begin{document}

\begin{frontmatter}

\title{The Gilmore-NASG model to predict single-bubble cavitation\\ in compressible liquids}

\author{Fabian Denner}
\ead{fabian.denner@ovgu.de}

\address{Chair of Mechanical Process Engineering, Otto-von-Guericke-Universit\"{a}t Magdeburg,\\ Universit\"atsplatz 2, 39106 Magdeburg, Germany}

\cortext[cor1]{Corresponding author: }

\begin{abstract}
The Gilmore model is combined with the Noble-Abel-stiffened-gas (NASG) equation of state to yield a simple model to predict the expansion and collapse of spherical bubbles based on real gas thermodynamics. 
\rev{The NASG equation of state resolves the temperature inaccuracy associated with the commonly employed Tait equation of state for liquids and, thus, can provide a consistent description of compressible and thermal effects of the bubble content and the surrounding liquid during cavitation. After a detailed derivation of the proposed Gilmore-NASG model,} the differences between the classical Gilmore-Tait model and the proposed model are highlighted with results of single-bubble cavitation related to bubble collapse and driven by an acoustic excitation in frequency and amplitude regimes relevant to sonoluminescence, high-intensity focused ultrasound \rev{and shock wave lithotripsy.
Especially for rapidly and violently collapsing bubbles, substantial differences in the bubble behaviour can be observed between the proposed Gilmore-NASG model and the classical Gilmore-Tait model.
The ability of the Gilmore-NASG model to simultaneously predict reliable pressure and temperature values in gas, vapour and liquid, makes the proposed model particularly attractive for sonochemistry and biomedical applications.}
\end{abstract}
\begin{keyword}
Cavitation \sep Bubble dynamics \sep Rayleigh-Plesset models \sep Real gas \\~\\
\textcopyright~2020. This manuscript version is made available under the CC-BY-NC-ND 4.0 license. \href{http://creativecommons.org/licenses/by-nc-nd/4.0/}{http://creativecommons.org/licenses/by-nc-nd/4.0/}
\end{keyword}
\end{frontmatter}

\section{Introduction}
The Rayleigh-Plesset (RP) model \citep{Plesset1949} and its various extensions are the workhorse of theoretical studies of bubble dynamics and cavitation. All RP models have in common that (i) the modelled bubble dynamics are spherical and that (ii) the compression and expansion of the gas or vapour inside the bubble is a polytropic process. Yet despite these seemingly limiting simplifications, RP models have been very successful in modelling a broad range of bubble dynamics and cavitation phenomena \citep{Lauterborn2010}.
Among the various extensions of the RP model to include the compressibility of the liquid or soft matter surrounding the bubble \citep{Keller1980, Trilling1952}, the model proposed by \citet{Gilmore1952} is generally considered one of the most widely applicable models \citep{Lauterborn2010, Fuster2011}.

\rev{The Gilmore model is traditionally founded on} the Tait equation of state (EOS), the polytropic form of the stiffened-gas EOS \citep{LeMetayer2004}, to describe the liquid. While the Tait EOS is able to represent the compressibility of common liquids, such as water, accurately, it fails to predict their temperature \citep{LeMetayer2016, Radulescu2020}, due to an unphysically large polytropic exponent and underpredicted heat capacity. This is a significant shortcoming for applications in which an accurate prediction of the bubble behaviour as well as the temperature \rev{and the pressure} of the liquid is critical, such as \rev{sonochemistry \citep{Yasui2005}}, sonocrystallisation \citep{Nalesso2019} or ultrasound-based medical treatments \citep{Wan2015, Izadifar2017}. 
 
\rev{During the expansion and collapse of a bubble, heat is exchanged between the gas (or vapour) inside the bubble and the surrounding liquid  as a result of thermal diffusion as well as evaporation and condensation \citep{Hauke2007}. The liquid, thus, regulates the temperature of the bubble content, whereby the large heat capacity of the liquid plays a dominant role \citep{Nigmatulin1981}. Representing the heat capacity of the liquid surrounding the bubble accurately, therefore, has a direct influence on the temperature and, in turn, on the pressure and chemical reactions inside the bubble.}

A conceptually simple model to predict the expansion and collapse of spherical bubbles based on a consistent EOS for both the gas or vapour inside the bubble as well as the liquid surrounding the bubble is not available to date. In order to remedy this shortcoming, a Gilmore model based on the Noble-Abel-stiffened-gas (NASG) EOS is proposed. The NASG EOS is able to describe both compressible and thermal effects in gases, liquids and their vapours reliably using a single EOS \citep{LeMetayer2016, Chiapolino2018, Radulescu2020}.

\section{Noble-Abel-stiffened-gas equation of state}
The NASG EOS \citep{LeMetayer2016} combines the stiffened-gas EOS and the Noble-Abel EOS to yield an unconditionally convex EOS that includes both molecular attraction and repulsion. 
The original idea for the NASG EOS goes back to \citet{Tammann1912}, as studiously pointed out by \citet{Radulescu2020}.

The NASG EOS is defined by its thermal and caloric equations of state, which are given as \citep{LeMetayer2016}
\begin{align}
p(v,T) &= \frac{(\gamma -1) \, c_v \, T}{v-b} - B  \label{eq:thermalEOS}\\
p(v,e) &= \frac{(\gamma -1) \, (e-q)}{v-b} - \gamma \, B  \label{eq:caloricEOS}
\end{align}
respectively,
where $p$ is the pressure, $T$ is the temperature, $e$ is the specific internal energy, $v$ is the specific volume, $b$ is the co-volume 
that represents the volume occupied by the individual molecules, $B$ is a pressure constant that models molecular attraction, $\gamma$ is the heat capacity ratio, $c_v$ is the heat capacity at constant volume and $q$ is a specific reference energy. For a fluid modelled by the NASG EOS, the 
speed of sound is 
\begin{equation}
c = \sqrt{\gamma \, \frac{p + B}{\rho \, (1-b \, \rho)}}, \label{eq:soundspeedNASG}
\end{equation}
where $\rho=1/v$ is the density, and the enthalpy is
\begin{equation}
h = \gamma \, c_v \, T + b \, p + q. \label{eq:hNASGfull}
\end{equation}
Along the isentrope, the NASG EOS yields \citep{Radulescu2020}
\begin{equation}
(p+B) \, (v-b)^\gamma = \text{const.}
\label{eq:isentrop}
\end{equation}

By replacing the heat capacity ratio $\gamma$ with the general polytropic exponent $\Gamma$, the isentropic relation given in Eq.~(\ref{eq:isentrop}) is readily turned into a polytropic NASG EOS, given as \citep{Denner2020a}
\begin{equation}
\frac{\rho}{1 - b \, \rho}  = K \, (p+B)^{\frac{1}{\Gamma}}, \label{eq:plytropicNASG}
\end{equation}
where $K$ is a constant representing the reference state, 
\begin{equation}
\rev{K = \frac{\rho_\text{ref}}{(p_\text{ref} + B)^{\frac{1}{\Gamma}} \, (1-b \, \rho_\text{ref})}, }
\end{equation}
\rev{with $p_\text{ref}$ and $\rho_\text{ref}$ the predefined reference pressure and reference density, respectively.}
The enthalpy of a polytropic NASG fluid follows by inserting Eq.~(\ref{eq:thermalEOS}) into Eq.~(\ref{eq:hNASGfull}) as
\begin{equation}
h = \frac{\Gamma}{\Gamma - 1} \frac{p+B}{\rho} -  \frac{\Gamma \, b}{\Gamma - 1} (p+B) + b \, p + q, \label{eq:enthalpyNASG}
\end{equation}
and the temperature of this polytropic process \rev{follows by inserting Eq.~(\ref{eq:thermalEOS}) into Eq.~(\ref{eq:isentrop})} as
\begin{equation}
T = T_\text{ref} \left(\frac{p+B}{p_\text{ref}+B}\right)^{\frac{\Gamma - 1}{\Gamma}}.
\label{eq:TNASG}
\end{equation}

Conveniently, the NASG EOS readily reduces to the Tait EOS ($b=0$, $B>0$), the ideal-gas EOS ($b=0$, $B=0$) or the Noble-Abel EOS ($b>0$, $B=0$), dependent on the chosen fluid parameters.

\section{Gilmore-NASG model}
The Gilmore model \citep{Gilmore1952} describes the temporal evolution of the bubble radius $R$ by a second-order ordinary differential equation, given as
\begin{equation}
\begin{split}
\left( 1 - \frac{\dot{R}}{C_\text{L}}\right) R \, \ddot{R} + \frac{3}{2} \left(1 - \frac{\dot{R}}{3 \, C_\text{L}} \right) \dot{R}^2 = \\ \left(1 + \frac{\dot{R}}{C_\text{L}}\right) H + \left( 1 - \frac{\dot{R}}{C_\text{L}}\right) R \, \frac{\dot{H}}{C_\text{L}} , \label{eq:gilmore}
\end{split}
\end{equation}
where $C_\text{L}$ is the speed of sound of the liquid at the bubble wall and $H$ is the difference between the enthalpy of the liquid at the bubble wall and at infinity.
Combining the Gilmore model, Eq.~(\ref{eq:gilmore}), with the NASG EOS, thus, requires suitable definitions for $C_\text{L}$ and $H$.

The density of the gas in the bubble follows from the conservation of mass as
\begin{equation}
\rho_\text{G} = \rho_\text{g,\text{ref}} \left(\frac{R_\text{ref}}{R} \right)^3,
\end{equation}
where $R_\text{ref}$ is the reference bubble radius and $\rho_\text{g,\text{ref}}$ is the associated gas density.
The pressure of the gas in the bubble is then given, following Eq.~(\ref{eq:plytropicNASG}), as
\begin{equation}
p_\text{G} = (p_\text{g,\text{ref}}+B_\text{g}) \left[\frac{\rho_\text{G} \, (1-b_\text{g} \, \rho_\text{g,\text{ref}})}{\rho_\text{g,\text{ref}} \, (1- b_\text{g} \, \rho_\text{G})} \right]^{\Gamma_\text{g}} - B_\text{g},
\end{equation}
where subscript $\text{g}$ denotes the properties of the gas and $p_\text{g,\text{ref}}$ is the predefined reference gas pressure associated with $R_\text{ref}$.
The pressure in the liquid at the bubble wall is defined as \citep{Lauterborn2010}
\begin{align}
p_\text{L}  = p_\text{G} -\frac{2 \, \sigma}{R} -  4 \, \mu_\ell \, \frac{\dot{R}}{R}, \label{eq:pL} 
\end{align}
where $\sigma$ is the surface tension coefficient and $\mu_\ell$ is the dynamic viscosity of the liquid.
The liquid speed of sound  at the bubble wall follows from Eq.~(\ref{eq:soundspeedNASG}) as
\begin{equation}
C_\text{L} = \sqrt{\Gamma_\ell \, \frac{p_\text{L} + B_\ell}{\rho_\text{L} \, (1 - b_\ell \, \rho_\text{L})}},
\end{equation}
where subscript $\ell$ denotes the properties of the liquid and
\begin{equation}
\rev{ \rho_\text{L} = \frac{K_\ell \, (p_\text{L} + B_\ell)^{\frac{1}{\Gamma_\ell}}}{1+ b_\ell \, K_\ell \, (p_\text{L} + B_\ell)^{\frac{1}{\Gamma_\ell}}}}
\end{equation}
is the density of the liquid at the bubble wall, \rev{with the constant representing the liquid reference state given as}
\begin{equation}
\rev{K_\ell = \frac{\rho_{\ell,\text{ref}}}{(p_{\ell,\text{ref}} + B_\ell)^{\frac{1}{\Gamma_\ell}} \, (1-b_\ell \, \rho_{\ell,\text{ref}})}, }
\end{equation}
\rev{where} $p_{\ell,\text{ref}}$ \rev{and $\rho_{\ell,\text{ref}}$ are} the predefined reference pressure \rev{and reference density} of the liquid, \rev{respectively}.
The enthalpy difference \rev{$H = h_\text{L}-h_\infty$} is given, based on Eq.~(\ref{eq:enthalpyNASG}), as
\begin{align}
H = \frac{\Gamma_\ell}{\Gamma_\ell - 1} \left(\frac{p_\text{L} + B_\ell}{\rho_\text{L}} - \frac{p_\infty+B_\ell}{\rho_\infty} \right) - b_\ell \, \frac{p_\text{L} - p_\infty}{\Gamma_\ell-1},
\end{align}
where
\begin{equation}
p_\infty = p_\text{L,0} + \rev{p_\text{a}} 
\label{eq:exc}
\end{equation}
is the pressure of the liquid at infinity, with \rev{$p_{\text{L},0}$ the ambient pressure in the liquid and $p_\text{a}$ the acoustic excitation pressure,}
and
\begin{equation}
\rev{ \rho_\infty = \frac{K_\ell \, (p_\infty + B_\ell)^{\frac{1}{\Gamma_\ell}}}{1+ b_\ell \, K_\ell \, (p_\infty + B_\ell)^{\frac{1}{\Gamma_\ell}}} }
\end{equation}
is the corresponding density at infinity.
The derivative of $H$ readily follows as
\begin{equation}
\begin{split}
\dot{H} &= \frac{\Gamma_\ell}{\Gamma_\ell-1} \left[\frac{p_\text{L} + B_\ell}{\rho_\text{L}} \left(\frac{\dot{p}_\text{L}}{p_\text{L} + B_\ell} - \frac{\dot{\rho}_\text{L}}{\rho_\text{L}} \right) \right. \\ &- \left.  \frac{p_\infty + B_\ell}{\rho_\infty} \left(\frac{\dot{p}_\infty}{p_\infty + B_\ell} - \frac{\dot{\rho}_\infty}{\rho_\infty} \right) \right] - b_\ell \, \frac{\dot{p}_\text{L} - \dot{p}_\infty}{\Gamma_\ell-1}, 
\end{split} \label{eq:dotH}
\end{equation}
where the derivatives of pressure are
\begin{align}
\dot{p}_\infty &= \dot{p}_\text{a}\\
\dot{p}_\text{L} &= \dot{p}_\text{G} + \frac{2 \, \sigma}{R^2} \dot{R} + 4 \, \mu_\ell \, \left(\frac{\dot{R}^2}{R^2} - \frac{\ddot{R}}{R}  \right) \label{eq:dotpL}\\
\dot{p}_\text{G} &=  \frac{\dot{\rho}_\text{G} \, \Gamma_\text{g} \, (p_\text{G} + B_\text{g})}{\rho_\text{G} \, (1-b_\text{g} \, \rho_\text{G})} 
%\dot{p}_\text{G} &=  \frac{\dot{\rho}_\text{G} \, \Gamma_\text{g} \, (p_\text{ref} + B_\text{g})}{\rho_\text{G} \, (1-b_\text{g} \, \rho_\text{G})}  \, \left[\frac{\rho_\text{G} \, (1-b_\text{g} \, \rho_\text{g,\text{ref}})}{\rho_\text{g,\text{ref}} \, (1- b_\text{g} \, \rho_\text{G})} \right]^{\Gamma_\text{g}}
\end{align}
and the derivatives of density are
\begin{align}
\dot{\rho}_\infty &= \frac{\dot{p}_\infty \, \rho_\infty}{\Gamma_\ell \, (p_\infty + B_\ell)} \, (1- b_\ell \, \rho_\infty)\\
\dot{\rho}_\text{L} &= \frac{\dot{p}_\text{L} \, \rho_\text{L}}{\Gamma_\ell \, (p_\text{L} + B_\ell)} \, (1- b_\ell \, \rho_\text{L}) \\
\dot{\rho}_\text{G} &= - 3 \, \rho_\text{G} \frac{\dot{R}}{R} .
\end{align}
\rev{Inserting the expressions for $\dot{\rho}_\infty$ and $\dot{\rho}_\text{L}$ into Eq.~(\ref{eq:dotH}) simplifies the  derivative of the enthalpy difference to
\begin{equation}
\dot{H} = \frac{\dot{p}_\text{L}}{\rho_\text{L}} - \frac{\dot{p}_\infty}{\rho_\infty}  . \label{eq:hdiff}
\end{equation}

Extending existing models based on the original formulation of \citet{Gilmore1952} to the improved formulation proposed above, merely requires to amend the definitions of density, pressure, speed of sound and enthalpy, as described in this section. The implementation of the Gilmore-NASG model is discussed in \ref{sec:implementation}.}

\section{Results}

To highlight the differences between the classical Gilmore-Tait model and the proposed Gilmore-NASG model, the results of four representative cavitation events of an air bubble in water are presented. \rev{The system of ordinary differential equations arising from Eq.~(\ref{eq:gilmore}), see Eqs.~(\ref{eq:ode1}) and (\ref{eq:ode2}) in \ref{sec:implementation}, is solved using a fourth-order Runge-Kutta method with adaptive time-stepping \citep{Dormand1980}.}
Air is described with $\Gamma_\text{g} = 1.4$, $B_\text{g} = 0$ and $\rho_\text{g,\text{ref}} = 1.2 \, \text{kg/m}^3$, and  $b_\text{g} = 0$ unless stated otherwise. Water has the properties $\Gamma_\ell = 1.19$, $B_\ell = 6.2178 \times 10^8 \, \text{Pa}$, $b_\ell = 6.7212 \times 10^{-4} \, \text{m}^3/\text{kg}$ and $\rho_{\ell,\text{ref}} = 997 \, \text{kg/m}^3$ \citep{Chandran2019} for the NASG EOS and $\Gamma_\ell = 7.15$, $B_\ell = 3.046 \times 10^8 \, \text{Pa}$ and $\rho_{\ell,\text{ref}}  = 997 \, \text{kg/m}^3$ for the Tait EOS. 
In all cases, the reference pressure is $p_{\text{g},\text{ref}} = p_{\ell,\text{ref}} = 10^5 \, \text{Pa}$. 
The reference temperature, which is not required to solve Eq.~(\ref{eq:gilmore}) and is defined only for post-processing, is $T_\text{ref} = 300 \, \text{K}$. \rev{Since thermal transport (advection, diffusion and radiation) and mass transfer are neglected, the liquid temperatures given below likely represent an upper limit.}

\begin{figure}
\includegraphics[scale=1]{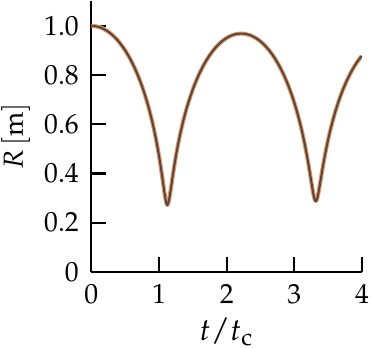} 
\hfill
\includegraphics[scale=1]{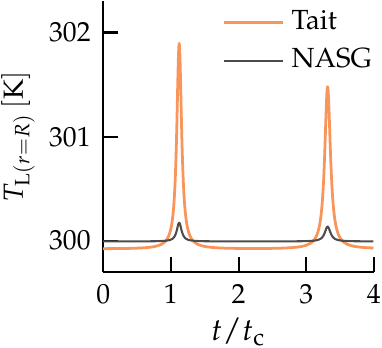}
\caption{Bubble radius $R$ and liquid temperature \rev{$T_{\text{L}(r=R)}$ at the bubble wall} as a function of dimensionless time $t/t_\text{c}$, with $t_\text{c} = 0.915 \, R_0 \, \sqrt{\rho_\infty/p_\infty}$ the Rayleigh collapse time, predicted by the Gilmore-Tait model and the Gilmore-NASG model for the Rayleigh collapse of a bubble with $R_0 = 1 \, \text{m}$ and $p_\infty/p_\text{G,0} = 10$.}
\label{fig:collapse}
\end{figure}

\begin{figure}
\includegraphics[scale=1]{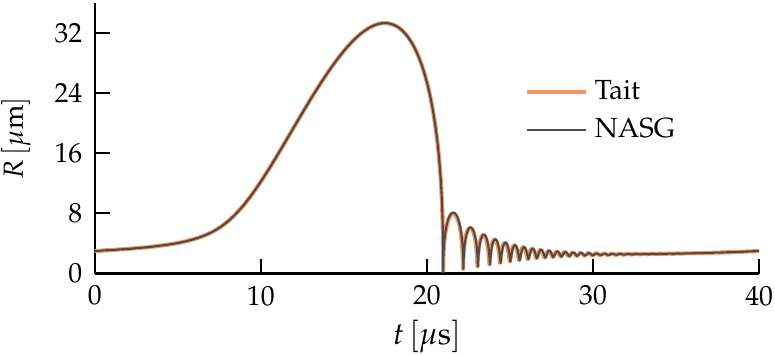}\\
\includegraphics[scale=1]{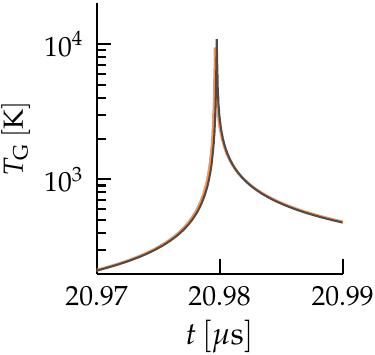} \hfill
\includegraphics[scale=1]{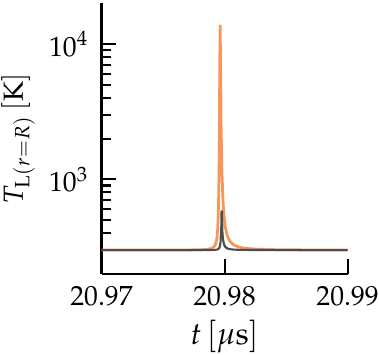}
\caption{Evolution of the bubble radius $R$, as well as the gas temperature $T_\text{G}$ and liquid temperature \rev{$T_{\text{L}(r=R)}$ at the bubble wall} during the first collapse, predicted by the Gilmore-Tait model and the Gilmore-NASG model for a bubble with $R_0=3 \, \mu \text{m}$ driven by an acoustic excitation with $\Delta p_\text{a}= 135 \, \text{kPa}$ and $f_\text{a} = 25 \, \text{kHz}$.}
\label{fig:sono}
\end{figure}

\begin{figure*}
\includegraphics[scale=1]{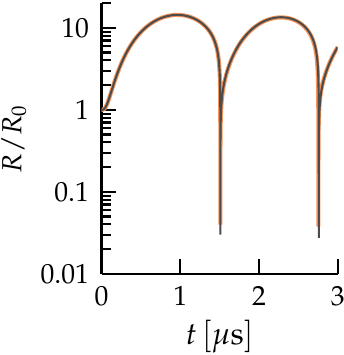}\hfill
\includegraphics[scale=1]{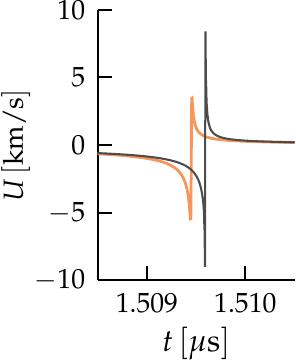}\hfill
\includegraphics[scale=1]{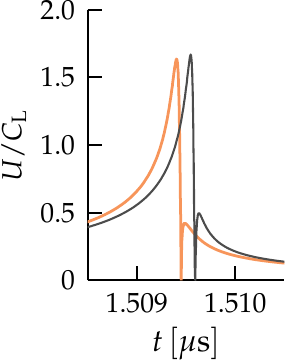}\hfill
\includegraphics[scale=1]{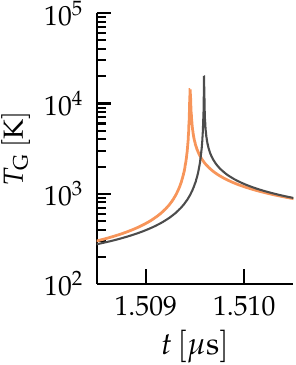}\hfill
\includegraphics[scale=1]{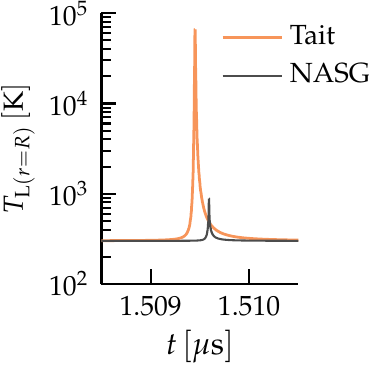}
\caption{Evolution of the bubble radius $R$, the velocity $U$ and Mach number $U/C_\text{L}$ of the bubble wall, and the temperature of the gas $T_\text{G}$ and the liquid \rev{$T_{\text{L}(r=R)}$ at the bubble wall}, predicted by the Gilmore-Tait model and the Gilmore-NASG model for a bubble with $R_0=1.25 \, \mu \text{m}$ driven by an acoustic excitation with  $\Delta p_\text{a} = 1.25 \, \text{MPa}$ and $f_\text{a} = 750 \, \text{kHz}$.}
\label{fig:HIFU}
\end{figure*}

\begin{figure*}
\includegraphics[scale=1]{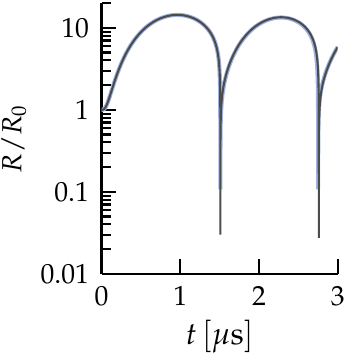}\hfill
\includegraphics[scale=1]{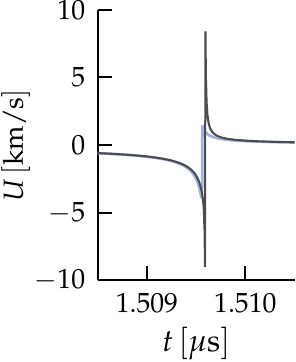}\hfill
\includegraphics[scale=1]{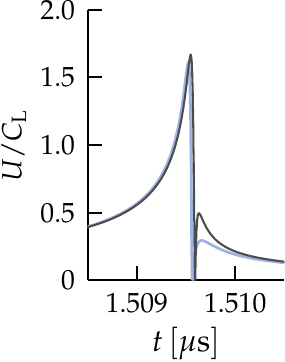}\hfill
\includegraphics[scale=1]{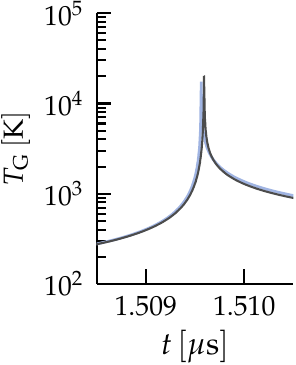}\hfill
\includegraphics[scale=1]{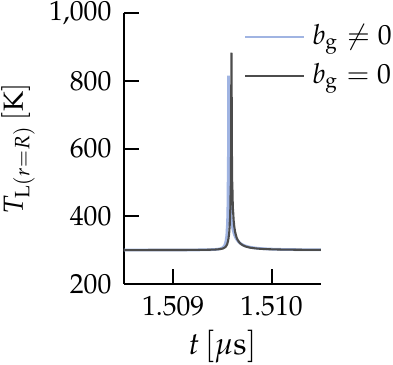}
\caption{Evolution of the bubble radius $R$, the velocity $U$ and Mach number $U/C_\text{L}$ of the bubble wall, and the temperature of the gas $T_\text{G}$ and the liquid \rev{$T_{\text{L}(r=R)}$ at the bubble wall}, predicted by the Gilmore-NASG model with $b_\text{g} = 0$ and $b_\text{g} = 10^{-3} \, \text{m}^3/\text{kg}$ for a bubble with $R_0=1.25 \, \mu \text{m}$ driven by an acoustic excitation with  $\Delta p_\text{a} = 1.25 \, \text{MPa}$ and $f_\text{a} = 750 \, \text{kHz}$.}
\label{fig:HIFU_b}
\end{figure*}

\rev{\subsection{Rayleigh collapse}}
First, a simple Rayleigh collapse of a bubble with \rev{initial radius} $R_0 = 1  \, \text{m}$ is considered, induced by an overpressure in the liquid at infinity of $p_\infty = 10^5 \, \text{Pa}$ against the initial gas pressure $p_\text{G,0}= 10^4 \, \text{Pa}$ in the bubble. Viscosity and surface tension are neglected. Fig.~\ref{fig:collapse} shows the evolution of the bubble radius $R$ and the temperature \rev{$T_{\text{L}(r=R)}$}, obtained via Eq.~(\ref{eq:TNASG}), of the liquid at the bubble wall.
While the bubble radius is in excellent agreement for both Gilmore models, the Gilmore-Tait model predicts a significantly higher temperature in the liquid. This is a manifestation of the unphysically large polytropic exponent of the Tait model ($\Gamma_\ell \approx 7$), chosen to approximate the compressibility of the liquid.

\rev{\subsection{Sonoluminescence}}
The cavitation of a bubble with \rev{initial radius} $R_0=3 \, \mu \text{m}$ driven by a \rev{sinusoidal} acoustic excitation \rev{defined as 
\begin{equation}
p_\text{a} = -\Delta p_\text{a} \, \sin(2 \pi f_\text{a} t),
\label{eq:sin}
\end{equation}
with} pressure amplitude $\Delta p_\text{a}= 135 \, \text{kPa}$ and frequency $f_\text{a} = 25 \, \text{kHz}$ is considered next, an acoustic regime relevant for sonoluminescence \citep{Brenner2002}. The bubble is initially in equilibrium, with $p_\text{G,0} = p_\text{L,0} + 2 \sigma/R_0$ and $p_\text{L,0} = 10^5 \, \text{Pa}$. The viscosity of the liquid is $\mu_\ell = 0.001 \, \text{Pa} \, \text{s} $ and the surface tension is $\sigma = 0.072 \, \text{N/m}$. The evolution of the bubble radius $R$ is shown in Fig.~\ref{fig:sono}, alongside the gas temperature $T_\text{G}$ and the liquid temperature \rev{$T_{\text{L}(r=R)}$ at the bubble wall} during the first collapse. \rev{While the temperature in the liquid at the bubble wall differs significantly for both models, the} 
evolution of the radius exhibits only very small differences between the two models. Interestingly, the temperature predicted in the liquid by the Gilmore-Tait model is higher than the corresponding gas temperature, a physically questionable result.

\rev{\subsection{High-intensity focused ultrasound}}
\label{sec:hifu}

The application of ultrasound-based diagnostic and therapy methods in biomedical applications requires an accurate prediction of the peak pressure amplitudes and the heat generated in the surrounding blood or tissue \citep{Wan2015,terHaar2011}. 
Fig.~\ref{fig:HIFU} shows the behaviour of a bubble with $R_0=1.25 \, \mu \text{m}$ and $\sigma = 0.072 \, \text{N/m}$ driven by a \rev{sinusoidal} acoustic excitation, \rev{Eq.~(\ref{eq:sin})}, with  $\Delta p_\text{a} = 1.25 \, \text{MPa}$ and $f_\text{a} = 750 \, \text{kHz}$, which is typical for high-intensity focused ultrasound treatments \citep{Coussios2008}. The viscosity of the liquid is $\mu_\ell = 0.001 \, \text{Pa} \, \text{s}$ and the bubble is initially in equilibrium, with $p_\text{L,0} = 10^5 \, \text{Pa}$.
The Gilmore-NASG model predicts a stronger collapse of the bubble than the Gilmore-Tait model, with a smaller minimum bubble radius and a considerably higher peak velocity of the bubble wall. However, the peak Mach number of the bubble wall, $M_\text{L} = U/C_\text{L}$, is similar with both models, since the Gilmore-NASG model also predicts a higher pressure and, consequently, a larger speed of sound of the liquid due to the stronger collapse of the bubble. In both cases the Mach number stays below $2.2$, the upper bound of validity previously proposed for the Gilmore model \citep{Gilmore1952}. The Gilmore-Tait model again predicts a peak temperature of the liquid at the bubble wall that exceeds the gas temperature.

\begin{figure*}
\includegraphics[scale=1]{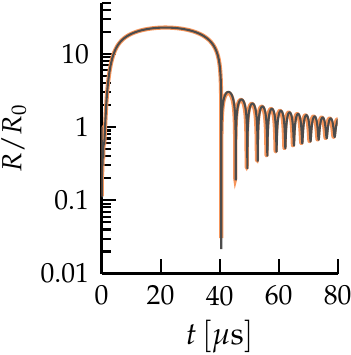}\hfill
\includegraphics[scale=1]{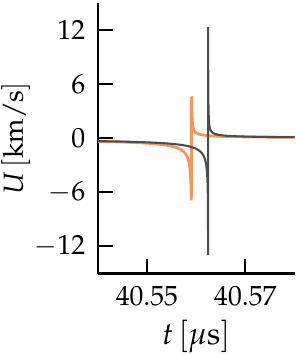}\hfill
\includegraphics[scale=1]{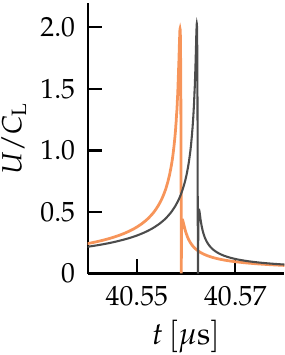}\hfill
\includegraphics[scale=1]{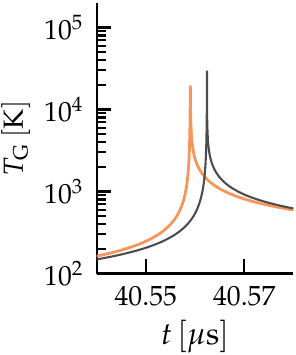}\hfill
\includegraphics[scale=1]{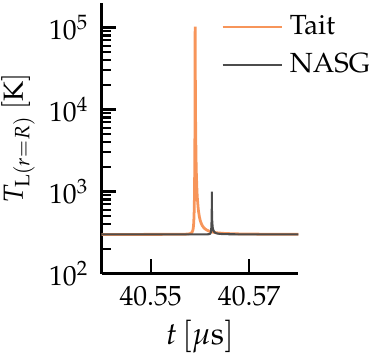}
\caption{\rev{Evolution of the bubble radius $R$, the velocity $U$ and Mach number $U/C_\text{L}$ of the bubble wall, and the temperature of the gas $T_\text{G}$ and the liquid $T_{\text{L}(r=R)}$ at the bubble wall, predicted by the Gilmore-Tait model and the Gilmore-NASG model for a bubble with $R_0=9 \, \mu \text{m}$ driven by a shock wave with  $\Delta p_\text{a} = 10 \, \text{MPa}$, $f_\text{a} = 83.3 \, \text{kHz}$ and $\alpha = 910 \, \text{kHz}$.}}
\label{fig:lithotripsy}
\end{figure*}

The cases presented above neglect the co-volume of the gas, $b_\text{g}$, in the NASG model. However, especially for inertial cavitation, where the gas in the bubble is compressed very strongly and rapidly, the volume occupied by the individual gas molecules becomes an important factor. Fig.~\ref{fig:HIFU_b} shows the same cavitation event as Fig.~\ref{fig:HIFU}, now using the Gilmore-NASG model with $b_\text{g} = 10^{-3} \, \text{m}^3/\text{kg}$, an approximate value typical for gases \citep{Longwell1958}, compared against $b_\text{g}=0$. The non-zero co-volume of the gas inhibits the collapse markedly, with a considerably reduced peak velocity of the bubble wall. The Mach number of the bubble as well as the temperatures of the gas and the liquid, however, do not change significantly. 

\rev{
\subsection{Shock wave lithotripsy}
Following the work of \citet{Church1989}, a shock-driven bubble collapse representative of shock wave lithotripsy treatments is considered. The bubble collapse is driven by a shock wave, defined as \citep{Church1989}
\begin{equation}
p_\text{a} = 2 \, \Delta p_\text{a} \, e^{-\alpha  t} \cos \left(2 \pi f_\text{a} t + \frac{\pi}{3}\right), 
\end{equation}
with $f_\text{a} = 83.3\, \text{kHz}$, $\Delta p_\text{a} = 10 \, \text{MPa}$ and the decay constant $\alpha = 910  \, \text{kHz}$. The bubble has an initial radius of $R_0=9 \, \mu \text{m}$ and is initially in equilibrium, with $p_\text{L,0} = 10^5 \, \text{Pa}$. The liquid has a viscosity of $\mu_\ell = 0.001 \, \text{Pa} \, \text{s} $ and the surface tension is $\sigma = 0.072 \, \text{N/m}$. The bubble behaviour, shown in Fig.~\ref{fig:lithotripsy}, exhibits similar differences between the classical Gilmore-Tait model and the proposed Gilmore-NASG model as the bubble considered in Section \ref{sec:hifu}; the peak velocity of the bubble wall predicted by the Gilmore-NASG model is considerably higher and the Gilmore-Tait model yields a higher peak temperature in the liquid than in the gas.}

\section{Conclusions}
A new model for the prediction of single-bubble cavitation in compressible liquids has been presented, by combining the Gilmore model \citep{Gilmore1952} and the Noble-Abel-stiffened-gas (NASG) equation of state \citep{LeMetayer2016}. The NASG equation of state provides a consistent description of compressible and thermal effects of both gases and liquids, resolving the temperature inaccuracy associated with the commonly used Tait equation of state. 
\rev{Even without considering thermal diffusion and mass transfer, which both play an important role in the dynamic behaviour of cavitation bubbles \citep{Hauke2007} but have not been considered in the presentation of the proposed model, significant differences in the bubble behaviour can be observed between the proposed Gilmore-NASG model and the classical Gilmore-Tait model, especially for rapidly and violently collapsing bubbles.
For the bubbles in the excitation regimes representative of high-intensity focused ultrasound and shock wave lithotripsy treatments, the velocity of the bubble wall  predicted by the Gilmore-NASG model is approximately twice as high as the velocity predicted by the Gilmore-Tait model, a difference that may be important for clinical safety considerations of such treatments \citep{Izadifar2017}.}

\rev{The ability of the Gilmore-NASG model to predict pressure and temperature values in gas, vapour and liquid simultaneously, makes the proposed model particularly attractive for sonochemistry and biomedical applications.
While an accurate description of evaporation and condensation together with a consistent model of the vapour inside the bubble are key to predict and understand chemical reactions occurring inside the bubble \citep{Storey2000, Xu2003, An2005}, the temperature distribution and the accumulation of heat in the liquid are primary concerns with respect to the efficacy and safety of medical treatments \citep{Wan2015, terHaar2011, Izadifar2017}.
All these phenomena necessitate an accurate temperature prediction in the liquid.}
The Gilmore-NASG model can, \rev{therefore,} serve as the foundation for future model developments, e.g.~for supercritical fluids \citep{Chiapolino2018} \rev{in sonochemistry applications}, 
and studies related to cavitation events in which an accurate knowledge and consistent definition of pressure and temperature of the liquid are critical, such as sonocrystallisation \citep{Nalesso2019} and medical ultrasound applications \citep{Wan2015}.
 
\begin{samepage}
\section*{Acknowledgements}
\noindent This research was funded by the Deutsche Forschungsgemeinschaft (DFG, German Research Foundation), grant number 441063377. 
\end{samepage}

\appendix

\rev{\section{Implementation of the Gilmore-NASG model}
\label{sec:implementation}}

\rev{Implementing a Gilmore model to predict the behaviour of a bubble in a viscous fluid  requires to rearrange Eq.~(\ref{eq:gilmore}), since the derivative of the liquid pressure at the bubble wall, $\dot{p}_\text{L}$, and, in turn, the derivative of the enthalpy difference, $\dot{H}$, are a function of the acceleration of the bubble wall, $\ddot{R}$, which is the primary solution variable. First,  defining the coefficient 
\begin{equation}
\mathcal{A} = \left( 1 - \frac{\dot{R}}{C_\text{L}}\right) R,
\end{equation}
inserting it in Eq.~(\ref{eq:gilmore}) and rearranging for $\ddot{R}$ leads to
\begin{equation}
\ddot{R} = \dfrac{\left(1 + \dfrac{\dot{R}}{C_\text{L}} \right) H - \dfrac{3}{2} \left(1 - \dfrac{\dot{R}}{3 \, C_\text{L}} \right) \dot{R}^2}{\mathcal{A}} + \dfrac{\dot{H}}{C_\text{L}}.
\label{eq:gilmore2}
\end{equation}
The derivative of the liquid pressure at the bubble wall, $\dot{p}_\text{L}$, defined in Eq.~(\ref{eq:dotpL}), is split into an explicitly treated part and an implicitly treated part, given as
\begin{equation}
\dot{p}_\text{L} = \dot{p}_\text{L,e} + \dot{p}_\text{L,i},
\end{equation}
with the explicitly treated part defined as
\begin{equation}
\dot{p}_\text{L,e} = \dot{p}_\text{G} + \frac{2 \, \sigma}{R^2} \dot{R} + 4 \, \mu_\ell \,  \frac{\dot{R}^2}{R^2}
\end{equation}
and the implicitly treated part is constituted by the term including $\ddot{R}$ and defined as
\begin{equation}
\dot{p}_\text{L,i} = - 4 \, \mu_\ell \, \frac{\ddot{R}}{R} .
\end{equation}
Splitting the enthalpy derivative, see Eq.~(\ref{eq:hdiff}), into explicitly and implicitly treated parts in a similar fashion, follows as
\begin{equation}
\dot{H} = \dot{H}_\text{e} + \dot{H}_\text{i}
\end{equation}
with
\begin{equation}
\dot{H}_\text{e} =  \frac{\dot{p}_\text{L,e}}{\rho_\text{L}} - \frac{\dot{p}_\infty}{\rho_\infty} 
\label{eq:hdiffE}
\end{equation}
and
\begin{equation}
\dot{H}_\text{i} =  \frac{\dot{p}_\text{L,i}}{\rho_\text{L}} = -4 \, \frac{\mu_\ell \,  \ddot{R}}{\rho_\text{L} \, R}.
 \label{eq:hdiffI}
\end{equation}
Inserting Eqs.~(\ref{eq:hdiffE}) and (\ref{eq:hdiffI}) into Eq.~(\ref{eq:gilmore2}) and rearranging for $\ddot{R}$ yields
\begin{equation}
\ddot{R} = \frac{\dfrac{\left(1 + \dfrac{\dot{R}}{C_\text{L}} \right) H - \dfrac{3}{2} \left(1 - \dfrac{\dot{R}}{3 \, C_\text{L}} \right) \dot{R}^2}{\mathcal{A}} + \dfrac{\dot{H}_\text{e}}{C_\text{L}}}{\mathcal{B}} \label{eq:gilmorefinal}
\end{equation}
with
\begin{equation}
\mathcal{B} = 1 + 4 \, \frac{\mu_\ell}{\rho_\text{L}\, R \, C_\text{L}} .
\end{equation}
A system of two first-order ordinary differential equations can then be readily defined based on Eq.~(\ref{eq:gilmorefinal}) as
\begin{align}
\dot{R} &= U \label{eq:ode1} \\
\dot{U} &= \frac{\dfrac{\left(1 + \dfrac{U}{C_\text{L}} \right) H - \dfrac{3}{2} \left(1 - \dfrac{U}{3 \, C_\text{L}} \right) U^2}{\mathcal{A}} + \dfrac{\dot{H}_\text{e}}{C_\text{L}}}{\mathcal{B}}, \label{eq:ode2}
\end{align}
which may be solved with any common ODE solver.
}

\end{document}